\documentclass[twocolumn,showpacs,preprintnumbers,amsmath,amssymb]{revtex4-1}
\usepackage{bm}
\usepackage{graphicx}
\begin{document}
\title{ Comment on $\frac{K}{\pi}$ Fluctuations at Relativistic Energies} 
\author{ 
H. Zheng$^a$ and A. Bonasera$^{a-b}$.
}
\affiliation{
a)Cyclotron Institute, Texas A\&M, College Station, 77843, TX, USA;\\
b)Laboratori Nazionali del Sud, INFN, via Santa Sofia, 62, 95123 Catania, Italy.
}

\begin{abstract}
We discuss the $\frac{K}{\pi}$  fluctuations in relativistic heavy ion collisions reported by the STAR collaboration, PRL${\bf 103}$,092301(2009), also discussed by the NA49 collaboration at SPS-CERN.  We show that these fluctuations
could be explained simply by imposing strangeness and charge conservations.  
\end{abstract}
\maketitle
Dynamic fluctuations of strange particles production have been proposed as a possible signature of a phase transition from Nuclear Matter (NM) to a Quark Gluon Plasma (QGP)\cite{star,na49}. ' Dynamical' quantities
$\nu_{dynK\pi}$ and $\sigma_{dyn}$ have been defined from the ratios of charged kaons to pions and those quantities become zero for Poissonian fluctuations. The two quantities are related and they are discussed at length in 
\cite{star,na49} and references therein.  In particular when $\sigma_{dyn}$ is plotted as function of the center of mass energy a different trend is observed in the STAR data than in the NA49 data, even though the points closest in energy
between the two experiments agree within errors, see figure 1.   
 A large literature is being produced to explain the dynamical fluctuations of these quantities and other particle ratios are being investigated as well\cite{star,na49}.  
Before getting into exciting physics one should be sure that trivial effects and/or autocorrelations are not playing a role in the observed quantities.  For instance the centrality is defined by looking at the particle multiplicities in a given rapidity or transverse energy bin.
Those particles are usually mostly pions that later are used to calculate the 'dynamical' fluctuations, thus the pion fluctuations might be reduced since they are the same particles used to determine centrality \cite{lat}. Kaons are also included to determine the centrality dependence, but are present in  much smaller
numbers than pions, so any possible autocorrelation is less important.  As a result dynamical fluctuations are dominated by the kaon multiplicities.  However, kaon production is subject to strangeness (and charge) conservation, in particular, on an event by event basis the net strangeness production
must be zero, which in turn is equivalent to saying that the number of $K^+$ and $K^-$ production is the same.  A strict equivalence is of course not achieved since other strange particles are produced in the reactions,  for instance the ratio at $\sqrt s=200GeV$ of  $\frac{K^-}{K^+}\approx 0.9$.
In this comment we show that the results reported in \cite{star,na49} can be simply explained assuming a Poissonian distribution for each particle type and generating 'events' from such a distribution.  Clearly for 'pure' Poissonian distributions zero dynamical fluctuations are obtained.
We impose that the number of  $K^+$ and $K^-$ differs at most 1-3 units.  The average numbers which enter the Poissonian distributions are taken from the data at each beam energy.  Imposing strangeness (and charge) conservation we obtain the 'dynamical' fluctuations plotted in figure 1, which agree rather well
with experimental data. 

We conclude that the observed 'dynamical' fluctuations are a consequence of strangeness and charge conservation.  A possible phase transition should give higher fluctuations which should dominate the $\nu_{dynK\pi}$ or $\sigma_{dyn}$ variables.
This might imply that there is no phase transition in the energy region investigated or possible autocorrelations might decrease the fluctuations.  Another possibility could be that the expansion during the collision of the system is too fast
and there is no time to develop critical fluctuations.  If the latter scenario is valid then it would be more suitable to look for other quantities that give a signature of the phase transition in a dynamical finite system.  

\begin{figure}[ht]
\centerline{\hspace{-0.4in}
\includegraphics[width=3.in,height=2.in,angle=0]{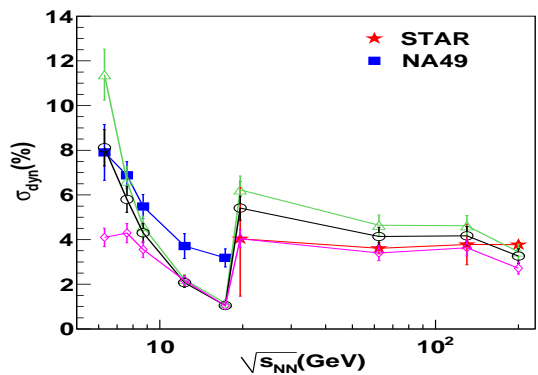}}
\vspace{0.01in} \caption{$\sigma_{dyn}$ as function of the center of mass energy.  Data from \cite{star,na49}.  Calculations are given by the open symbols,
$|N_{k^+}-N_{k^-}|$ $\le1$(triangles), $2$(circles) or $3$(diamonds). The stronger the constraint the larger the fluctuations are. The best agreement with data is obtained when the charged kaons difference is at most two.}
\label{fig1}
\end{figure}









\end{document}